\documentstyle[epsfig,aps,prb,twocolumn]{revtex} 
\begin{document}
\tightenlines

\wideabs{

\title{High-Pressure Amorphous Nitrogen} 
\author{Eugene Gregoryanz, Alexander F. Goncharov, Russell J. Hemley 
and Ho-kwang Mao}
\address{Geophysical Laboratory and Center for High Pressure
Research, Carnegie Institution of Washington, \\ 
5251 Broad Branch Road NW, Washington D.C. 20015 U.S.A \\ } 
\maketitle

\begin{abstract}

The phase diagram and stability limits of diatomic solid nitrogen have
been explored in a wide pressure--temperature range by several 
optical spectroscopic techniques. A newly characterized narrow-gap 
semiconducting phase $\eta$ has been found  to  exist in  a range of 
80--270 GPa  and 10--510  K. The vibrational  and optical  properties 
of the $\eta$ phase  produced under these conditions indicate that  
it is largely amorphous and back transforms to a new molecular phase. 
The band gap of the  $\eta$ phase is  found  to decrease  with pressure  
indicating possible metallization by band overlap above 280 GPa.
\end{abstract} 
}

Despite  early theoretical  predictions for a transformation  of  
nitrogen   to a  monoatomic  state, \cite{mcmahan,martin,mailhiot}  reliable
experimental  evidence became available  only  quite recently.
\cite{pres,los}  Optical spectroscopy, visual  observations and
electrical conductivity measurements showed that the material 
transforms to a semiconducting non-diatomic phase at 150 GPa 
(190 GPa at 80 K).\cite{pres,los}
The transition to the nonmolecular state was predicted to
be hindered  by a  large energy barrier  and accompanied by  a 
large volume discontinuity  and hysteresis. \cite{martin,mailhiot} 
The  latter was confirmed by visual observations which  
indicated that  the  high-pressure phase  can  be preserved  
to  17  GPa at  low temperatures. \cite{los}

Characterization  of the high-pressure  phase (called  $\eta$ here)
still remains  an important  issue because of  the lack  of structural
studies   and   systematic  spectroscopic   data   at  different   P--T
conditions. Optical absorption spectra \cite{pres} reveal the presence
of a low-frequency logarithmic  Urbach tail \cite{urbach} and a higher
energy region,  which obeys the  empirical Tauc law.  \cite{tauc} This
type of  absorption edge is typical for  amorphous semiconductors (see
Ref.  \onlinecite{mott}).    Although  being  quite  diagnostic,  this
observation  still  requires  confirmation,  as  a  highly  disordered
high-pressure  structure   is  consistent  with  the   nature  of  the
transformation; i.e. a large volume change  at a reconstructive
phase   transition  can   cause  large   shear  stresses   because  of
inhomogeneous   nucleation    of   the   high-pressure    phase   (see
Ref. \onlinecite{jeanloz}).  Moreover,  experiments demonstrate
that two  phases coexist  in a  wide pressure range  (at 300  K), 
\cite{pres} thus making the characterization of the $\eta$ phase  
even  more difficult. Here we present new optical data over a wide 
P--T range indicating that high-pressure non-molecular phase is a 
largely amorphous, narrow gap semiconductor to at least 268 GPa. 
We examine the stability limits of the diatomic molecular state and 
present evidence for new transformations, including 
metallization by band overlap above 280 GPa.

Four experiments were performed  at room temperature  with the 
maximum pressures varied from 180 to 268  GPa.  Above 200 GPa 
pressure was determined using tunable  red lines of Ti:sapphire
laser combined with time resolving technique (Fig. 1).
For low-temperature  
measurements we used a continuous-flow  He  cryostat,  which 
allowed infrared and {\it in situ} Raman/fluorescence measurements. 
\cite{gonchy}  
High-temperature Raman and visible transmission
measurements were performed with  an externally  heated  cell.
\cite{fei} In  this case, infrared measurements were  done on samples
quenched to room temperature.

Fig.  2  shows  representative  IR and  visible  transmission  spectra
demonstrating the effect of  temperature on the semiconducting optical
edge  characteristic  of the $\eta$  phase.  The  spectra  presented
correspond to the conditions when no molecular phase is present in the
sample as determined from vibrational spectroscopy (see below).
No variation of the shape and position of the band gap can be detected
from  transmission  spectra  at  different temperatures  and  constant
pressure of 200 GPa.  Fig. 2b shows that the low-energy portion of the
spectra  plotted in logarithmic  scale (Urbach  plot) have  a constant
slope ($\Gamma$) in  a 10--200 K  range. This also agrees with 300 K 
data, \cite{pres}  thus  showing  that   $\Gamma$  is  not temperature
dependent. Similar spectra have been reported for amorphous phosphorus 
at zero pressure. \cite{pilione} This is typical for solid amorphous 
semiconductors, \cite{mott} because the random microfield is caused 
by static  disorder  in  the  system  as opposed to crystalline 
materials \cite{urbach}  where the vibrations generate a 
temperature-dependent  dynamical disorder.

Determination of  the band gap  from our data  is a  complicated issue,
because there is no characteristic feature of the spectra which can be
associated with  the band gap (e.g., Ref. \onlinecite{knief}).   This is
especially important  for our measurements, since  we essentially deal
with  samples of  various thickness (which is  a function  of anvil
geometry and pressure).  As the  result,  visual observations of
the  sample above  230  GPa showed  that  it is  red  or yellowish  in
transmission  and black in  reflection, which  is consistent  with the
semiconducting  state.  The color  of  the  sample  (compare with  the
observations of dark  nitrogen in Refs. \onlinecite{pres,los,reichlin,bell})
may be explained by its thickness  (of the order of 1 $\mu$m) compared
to the  samples brought to  150 GPa (up  to 5 $\mu$m). At  the highest
pressure (268  GPa), visible  transmission spectra clearly  show the
presence  of  the   fundamental  absorption  edge  characteristic of
semiconductors (Fig.  2a). This result  is in agreement  with direct
electrical measurements performed to 240 GPa. \cite{los}

The high-energy absorption  edge, which can be observed  in this case,
corresponds  to  electronic  transitions  between  extended  states
(unlike Urbach  absorption, which is presumably  caused by transitions
from localized to extended states). Extrapolation of the absorption
spectra  plotted as (h$\nu  \alpha)^{0.5}$  versus h$\nu$
gives the value of optical  gap. \cite{tauc} These
values  at  different  pressures are shown in
Fig. 3. Note  that data from different experiments  agree, despite the
different sample  thickness and the fact  that some of  the data are
taken on pressure release in a metastable pressure region (see below).
We observed  a monotonic red-shift of  the band gap  with pressure (see
also Fig. 2a). The pressure  dependence of the band gap is sublinear
mainly due to contribution  from the points obtained on decompression.
The extrapolation  of the band gap values gives metallization  at pressures
slightly above 300  GPa.  Linear extrapolation of this  curve to higher
pressures (not taking into account points obtained upon decompression)
gives a value  of 280 GPa. 

We now present temperature measurements of the vibrational properties of the
$\eta$ phase. Type II diamonds  were used for  mid-IR measurements 
to  avoid interference with the  characteristic absorption of the  
sample. The representative absorption spectra at different temperatures 
(see Fig. 2) clearly show the presence of a broad 1700 cm$^{-1}$ IR 
band (compare with Ref. \onlinecite{pres}). Its presence was  also observed 
in the sample  heated to 495 K  at 117 GPa (see below).  The position 
of the  band and its damping  (if fitted as one band) does not depend 
on pressure and temperature within the error bars.

The  Raman spectrum of the $\eta$ phase obtained on heating
(see below) does not  show any  trace of
the molecular phase (see Fig. 4b).  Careful
examination of the  spectrum in this case showed a  weak broad band at
640 cm$^{-1}$  in agreement  with the data  of Ref. \onlinecite{pres}  and a
shoulder near 1750  cm$^{-1}$ (both indicated with arrows in Fig. 3b). 
The later may indicate  the presence of
the second Raman  peak close to the position of  the observed IR band,
but can also  be due to Raman in the stressed  diamond. The broad two-peak
structure of  the phonon spectrum  of the material is  consistent with
its  amorphous  nonmolecular  nature.  For  an  amorphous  state,  the
spectrum observed  would closely resemble  a density of  phonon states
\cite{brodski}  with the  maxima  corresponding roughly to  the zone  
boundary acoustic and  optic vibrations of  an underlying structure.
\cite{structure}   The  only   lattice   dynamics  calculations   for
hypothetical   high-pressure  crystalline   phases  of   nitrogen  are
available  for   the  cubic  gauche  phase, \cite{barb}  and
calculated  phonon  frequencies  are  in  a good  agreement  with  our
measurements.  The vibrational  spectroscopy and  band gap
structure indicate the absence of a long-range order. The material 
can still possess some short-range order, for example related to 
pyramidal coordination of nitrogen atoms. The absence of the long-range order
can be due to the structural flexibility because each atom forms
bonds with only three other atoms  out of 6 nearest 
neighbors. \cite{structure}  

We  probed  the forward  and  reverse  transformation  of the molecular  to
{$\eta$}  phase in  different  regions  of P-T  space.  We used  IR
transmission spectra as diagnostics of the degree of transformation to
the  nonmolecular phase.  The  absence of  IR  bands corresponding  to
vibrons and lattice modes of  the molecular phase \cite{pres} was used
as  a  criteria. Since  both the  molecular  and  nonmolecular phases  are
transparent in the  mid-IR, the amount of the  phase present is simply
{\em  proportional}   to  the   amplitude  of  the   corresponding  IR
peaks. This is unlike the situation with Raman spectra which are 
attenuated by absorption the {$\eta$} phase. 
We examined  the transformation  at 205 K  and elevated  pressures and
found that it starts at 155  GPa and completes at 185 GPa. 
This is shifted to higher pressures compared to our 300 K data
\cite{pres}   and is  in   agreement   with  the   trend   reported   in
Ref. \onlinecite{los}. The  sample has been  cooled down to  10 K at 200  GPa and
warmed up after subsequent release of pressure at 130-150 GPa.  IR and
visible  transmission spectra  and  Raman spectra  clearly showed  the
persistence  of {$\eta$} phase  without any  reverse transformation
down to 120 GPa. At this point  the pressure dropped to 87 GPa and the
sample transformed instantaneously back to a transparent phase (called
$\zeta^{'}$ here). The molecular nature  of this phase is confirmed by
its Raman  spectrum (Fig. 4a) although  the positions of  the vibron 
lines do not  correspond  to  those observed  on pressure 
increase  (Fig.  4c). This means that amorphous phase back transforms 
to a molecular phase which  differs from the one observed on upstroke. 
On further release of pressure (to 60  GPa) we observe the Raman 
spectra which are similar to those of $\zeta$ or $\epsilon$ phases
in positions and intensities of vibron peaks. The quality of the 
spectra (on pressure release, the sample thins down leading
to the considerable loss of intensity) does not allow to 
establish the presence/absence of the weaker vibron modes  
characteristic of $\zeta$ phase and unambiguously  determine
whether $\zeta^{'}$ phase back transforms to $\zeta$     
or $\epsilon$ phase.  

In the heating experiment we first exposed  the sample to 495  K at 117
GPa.   The  effect of temperature caused a  gradual transformation 
(starting at  450  K) similar to that observed at 300 and 200 K. 
The comparison  of  Raman  modes revealed  more  than a  10-fold
decrease of intensity  in the Raman vibrons and  no observable lattice
modes.  Quenching of the sample to room temperature did not change the
color and the visible  absorption spectra.  Surprisingly, the infrared
spectra  revealed  the   presence  of  molecular  vibrons,  indicating
an incomplete  transformation  (about  30\%  of  nonmolecular  phase
judging from the infrared activity).  During the second heating the sample
was  completely  transformed  to  the $\eta$  phase.  Then pressure 
was dropped  to 105  GPa at 460  K causing an instantaneous reverse
transformation to a transparent molecular phase. The spectral positions
of the bands  and their number do not correspond  to those observed at
this pressure on  compression but are similar to those obtained during the 
unloading at 300 K (see above).  Increasing  pressure  to 135  GPa at 510 K
drove the direct transformation into the $\eta$ phase again.

Fig. 5 summarizes our data for the phase diagram of nitrogen obtained in a
course  of  extensive  P--T  measurements. Substantial  hysteresis  is
observed for the transformation from  and back to the molecular phase,
so the  observed curves should be treated  as kinetic 
boundaries.  For the  direct transformation,  our data  are in  good
agreement  with  the results  of  visual  observations of Ref. \onlinecite{los}.  
Our high-temperature data show that  the hysteresis becomes quite 
small at  temperatures  above 500  K. There is  large  hysteresis 
at lower temperature such that the molecular  {$\zeta$} phase   
can be metastably  retained beyond  the $\zeta$--$\zeta^{'}$   
boundary  (above approximately  100 GPa; see also Ref. 
\onlinecite{los}).  Thus, the  observation of  another molecular phase  
($\zeta^{'}$)   in  this  P-T  conditions  means  that this 
phase is either kineticaly favored or thermodynamically stable with 
respect to the {$\zeta$} phase.  

If  the potential barrier  between two crystalline  phases is
high  (molecular dissociation  is required  in our  case), the
transition  may be preempted  by a  transformation to  
metastable  phase, which may  be amorphous. \cite{sharma}  
This defines an intrinsic stability limit (e.g., spinodal) 
for the diatomic molecular state of nitrogen. 
In view of the amorphous component of the higher 
pressure phase, the transition may be considered as a type of 
pressure-induced amorphization.  As such, the transformation boundary could 
track the metastable extension of the melting line of the molecular 
phase, and if so it should have a negative slope
(consistent with negative $\Delta$V and positive $\Delta$S for 
a transition to dense amorphous state. \cite{sharma})  
Alternatively, one can view this in terms of an intrinsic 
(elastic or dynamical) instability of the structure of the 
molecular solid.  In this sense, 
the behavior of the material parallels other amorphizing systems that undergo 
coordination changes (see Ref. \onlinecite{hemley}). 

The authors are grateful to  Y. Fei for the help with high-temperature
experiment. Special thanks to J. Badro and M. Somayazulu for 
their comments on high-pressure 
amorphization.  This work is supported by NSLS, NSF and DOE.

 

\begin{figure} 
\centerline{\epsfig{file=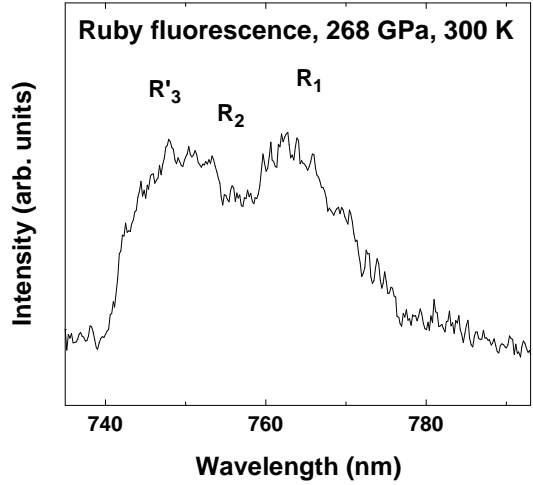,width=10cm}}
\caption{Ruby fluorescence spectrum at pressure 268 GPa and 300K.
Ruby was excited with 730 nm line of Ti:Sapphire laser.}
\label{fig1} 
\end{figure}

\begin{figure} 
\centerline{\epsfig{file=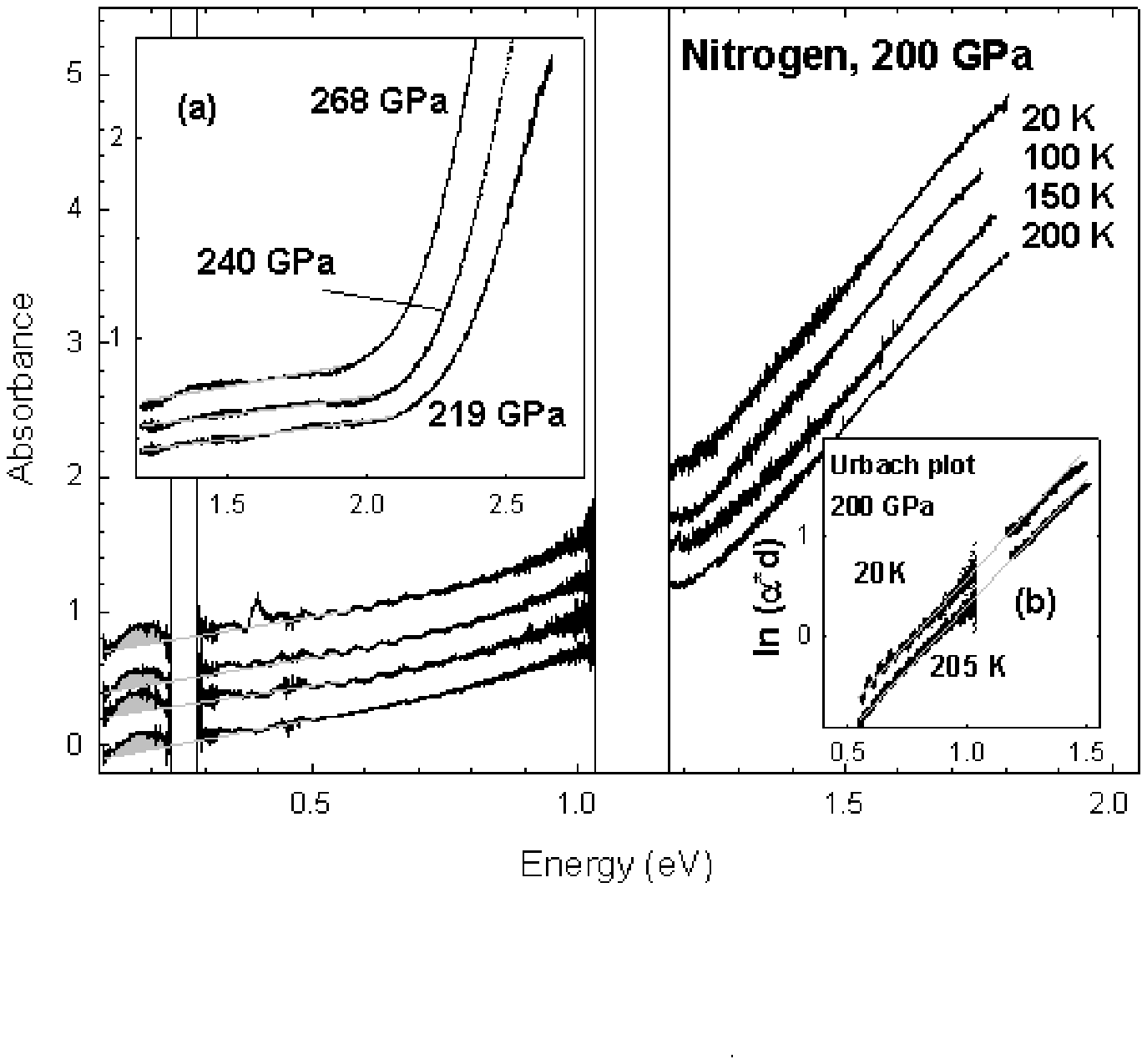,width=10cm}}
\caption{Transmission   spectra   of    N$_2$   as   a   function   of
temperature. Spectra  are shifted vertically  for the 
clarity. The characteristic peak of the {$\eta$} phase is shaded. Inset
(a) shows the  pressure dependence of the absorption  spectra of N$_2$
at very high pressures and  room temperature. Gray lines represent the
Tauc  fits  to the  spectra  in  an  appropriate spectral  range.  The
determination of the energy gap from these measurements is obscured by
additional losses  caused by a presence  of a fine ruby  powder in the
chamber.  The high-energy  absorption  edge is  most  probably due  to
stress    induced    absorption    of    diamond    anvils    \protect
\onlinecite{vohra}.  (b)  Urbach  plots   at  200  GPa  and  different
temperatures (shifted vertically). Gray lines are guides to the eye.}
\label{fig2} 
\end{figure}

\begin{figure} 
\centerline{\epsfig{file=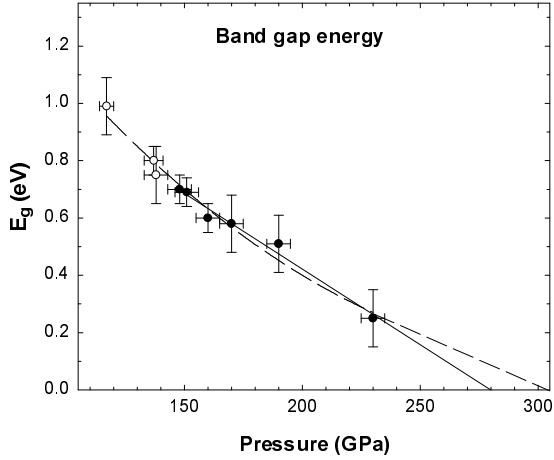,width=10cm}}
\caption{Band gap of the $\eta$  phase as a function
of pressure. Solid circles represent increasing pressure 
and open circles decreasing pressure. Linear and quadratic
extrapolation are shown in dashed lines.}
\label{fig3}
\end{figure}

\begin{figure} 
\centerline{\epsfig{file=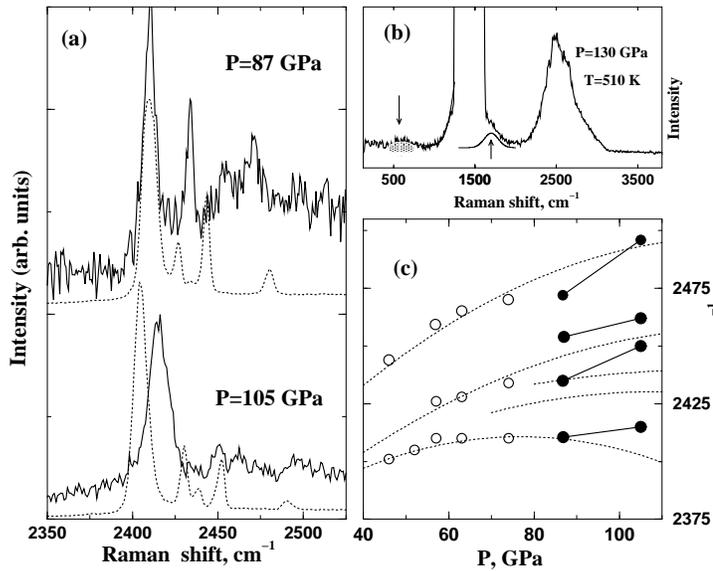,width=10cm}}
\caption{(a) Raman spectra in the $\zeta$ (dotted lines) 
and $\zeta^{'}$ (solid lines) phases. (b) Raman spectra 
of the $\eta$ phase (indicated with arrows); the fitted curve 
shows peak at 1700 cm$^{-1}$. (c) Raman shifts 
of molecular phases versus pressure. Dotted lines represent
shifts on compression. Solid lines and open circles
represent shifts on decompression from  the 
\protect $\eta$ phase.} 
\label{fig4}
\end{figure}

\begin{figure}
\centerline{\epsfig{file=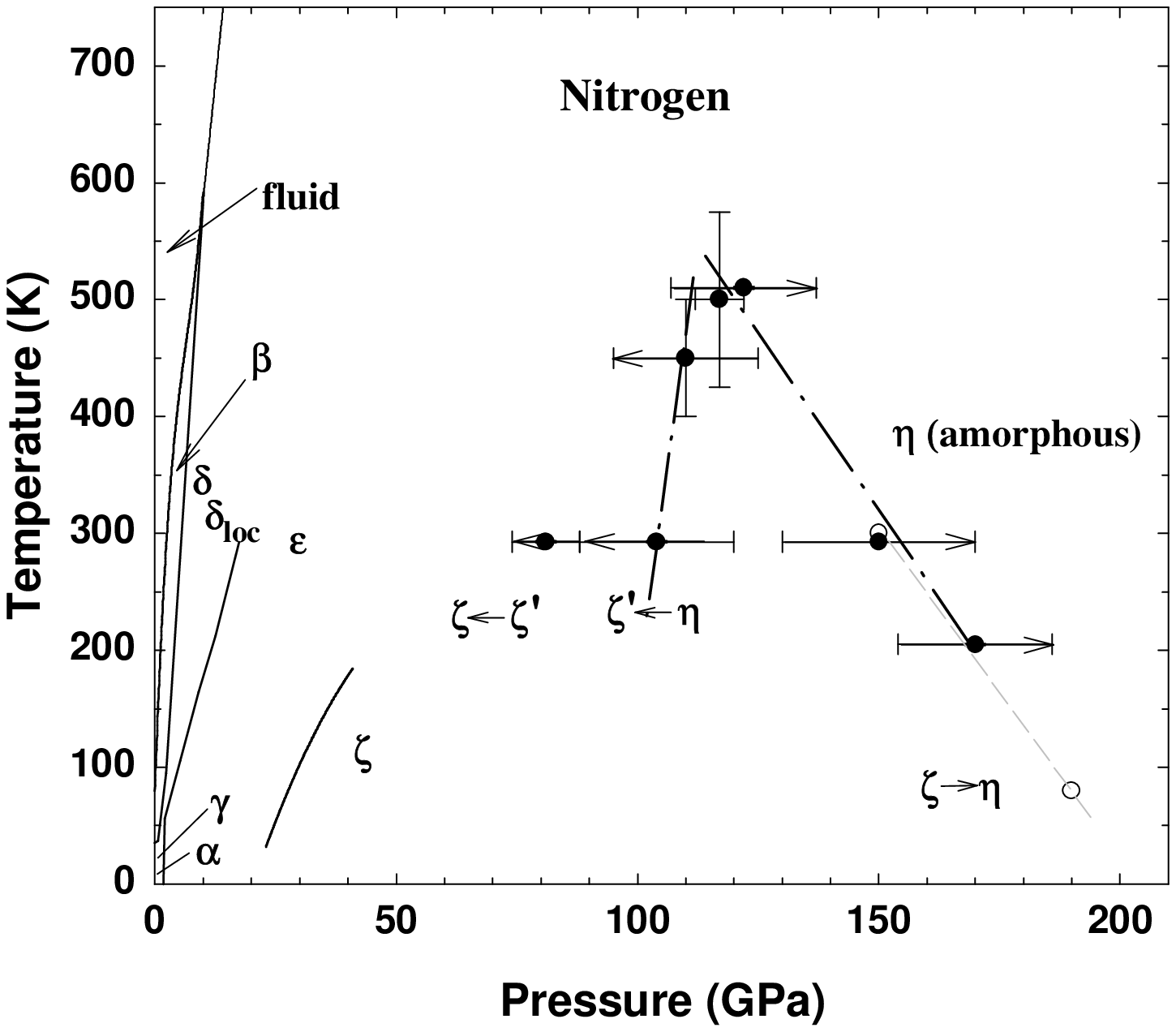,width=10cm}}
\caption{Phase diagram of nitrogen in a wide P--T  range.  Filled circles,
dashed-dotted lines and  arrows  are  the  data  from  this  work.  Open  circles 
and dashed line are from visual observations of Ref. \protect \onlinecite{los}. 
Arrows show the paths along which the
transformation  boundaries  were crossed.   The  length  of the  arrow
represents the  width of the two-phase region  or pressure uncertainty
(on a  pressure release). Phase  boundaries at low pressures  are from
Refs.   \protect \onlinecite{young,bini}.  The  phase boundaries
for $\alpha, \gamma$ and $\delta_{loc}$ phases are not shown.  
There  is evidence for further pressure-induced transformations of the high-pressure
molecular $\zeta$  phase but the products  are believed  to be  closely related.
\protect  \cite{reichlin,olijnyk99} Dotted line is extrapolation of 
the $\zeta$-$\eta$ transformation boundary.}
\label{fig5}
\end{figure}


\begin{references}
\bibitem{structure}  Theoretical  calculations \cite{mailhiot}  reveal
several candidates for the high-pressures phase. All three-coordinated
structures  can be  obtained by  distortion  of a  simple cubic  phase
(6-fold coordinated which  is unstable)  and are  related  to 
phonon instabilities  in
different high-symmetry points of the Brillouin zone. \cite{barb}
\end{references}
\end{document}